# Cosmological origin for cosmic rays above $10^{19}$eV

Eli Waxman

Institute for Advanced Study, Princeton, NJ 08540; e-mail: waxman@sns.ias.edu

To appear in the October 10 (1995) issue of the Ap. J. Letters

## ABSTRACT

The cosmic ray spectrum at $10^{19}$eV $- 10^{20}$eV, reported by the Fly's Eye and the AGASA experiments, is shown to be consistent with a cosmological distribution of sources of protons, with a power law generation spectrum $d \ln N / d \ln E = -2.3 \pm 0.5$ and energy production rate of $4.5 \pm 1.5 \times 10^{44}$erg Mpc$^{-3}$ yr$^{-1}$. The two events measured above $10^{20}$eV are not inconsistent with this model. Verifying the existence of a "black-body cutoff", currently observed with low significance, would require $\sim 30$ observation-years with existing experiments, but only $\sim 1$ year with the proposed $\sim 5000$ km$^2$ detectors. For a cosmological source distribution, no anisotropy is expected in the angular distribution of events with energies up to $\sim 5 \times 10^{19}$eV.

*Subject headings:* cosmic rays — acceleration of particles — gamma rays: bursts

## 1. Introduction

Recent results, reported by the Fly's Eye (Bird *et al.* 1993, 1994) and the AGASA (Hayashida *et al.* 1994, Yoshida *et al.* 1995) experiments, show two major features in the cosmic ray (CR) energy spectrum above $10^{17}$eV. A break in the shape of the spectrum is observed at $\sim 5 \times 10^{18}$eV; the CR composition also changes from predominantly heavy nuclei below the break to predominantly light nuclei above the break. Evidence for these features existed in the results of previous experiments with weaker statistics [see e.g. Watson (1991) for a review]. Coupled with the lack of observed anisotropy, these features strongly suggest that below $\sim 10^{19}$eV the cosmic rays are mostly heavy ions of Galactic origin, and that an extra-Galactic component of protons takes over above $\sim 10^{19}$eV. If the particles observed above $10^{19}$eV are indeed protons of extra-Galactic origin, then their interaction with the microwave background radiation (MBR) is expected to produce a "black-body cutoff" in the CR spectrum at $\sim 5 \times 10^{19}$eV (Greisen 1966, Zatsepin & Kuzmin 1966).

In this *Letter* we compare the CR spectrum above $10^{19}$eV, reported by the Fly's Eye and the AGASA experiments, with that expected from a homogeneous cosmological distribution of CR sources. We assume that each source generates high energy protons with a power law differential spectrum $dN/dE \propto E^{-\alpha}$. The expected spectrum is shown to be insensitive to source evolution with redshift and to the cosmological parameters, such as the Hubble constant and the average universe density. We find that current observations are consistent with a cosmological distribution of sources with $\alpha \approx 2$ and energy production rate of $\sim 4 \times 10^{44}$erg Mpc$^{-3}$ yr$^{-1}$ in the range $10^{19} - 10^{21}$eV. However, the statistical significance with which the black-body cutoff is observed in current data is small. We show that the signature of the interaction with the MBR should be sought in the energy range $2 \times 10^{19} - 10^{20}$eV rather than above $10^{20}$eV. At energies above $10^{20}$eV the spectrum is sensitive to inhomogeneities in the source distribution over scales $\sim 10$ Mpc. We also show that for a cosmological source distribution, no anisotropy is expected in the angular distribution of events with energies up to $\sim 5 \times 10^{19}$eV.

## 2. Method.

As they propagate, high energy protons lose energy due to the cosmological redshift and due to production of pions and $e^+e^-$ pairs in interactions with MBR photons. Here we approximate the energy loss due to scattering, which is a random process, as a continuous energy loss (CEL). In this approximation, the energy loss rate of a proton of given energy is taken to be the mean loss rate of an ensemble of protons of the same energy. The CEL approximation is excellent for pair production, for which the mean free path is small and the relative energy loss in a single scattering, of order $m_e/m_p \sim 10^{-3}$, is also small. For pion production, the average relative energy loss in a single collision is 0.13 at the threshold and rises to 0.5 at higher energy. However, fluctuations in proton energy due to this process are significant only for propagation distances smaller than 100 Mpc and proton energy $> 10^{20}$eV (Aharonian & Cronin 1994). The CEL approximation therefore gives accurate results for the flux below $10^{20}$eV. At higher energies the flux obtained by this approximation drops faster than that obtained when fluctuations in proton energy are taken into account. It has been shown (Berezinski, Grigor'eva & Zatsepin 1975, Berezinski & Grigor'eva 1988) that for a flat generation spectrum, $\alpha \leq 2.6$, the flux obtained using the CEL approximation is accurate to better than 10% up to $3 \times 10^{20}$eV, the highest energy at which events have been reported. We have confirmed this by comparing the Monte Carlo calculation of the spectrum obtained for cosmological sources with $\alpha = 2$, given in Yoshida & Teshima (1993), to our results. We find an agreement to better than 20% up to $2 \times 10^{20}$eV [For the comparison we have used the energy loss rate given by Fig. 3 of Yoshida & Teshima (1993). This rate is close to our approximation (described below) everywhere except near $5 \times 10^{19}$eV, where we find that Yoshida & Teshima under estimate the energy loss rate by $\sim 25\%$].

The energy loss rate due to cosmological redshift is $E^{-1} \mathrm{d}E/\mathrm{d}t = (1+z)^{-1}\mathrm{d}z/\mathrm{d}t$, where $z$ is the redshift. The energy loss rate due to pair or pion production is

$$\tau(E,z)^{-1} \equiv -\frac{1}{E}\frac{\mathrm{d}E}{\mathrm{d}t} = \frac{m_p^2 c^2 T(z)}{2\pi^2 \hbar^3 E^2} \int_{\epsilon_{th}}^{\infty} \mathrm{d}\epsilon \, \sigma(\epsilon)\eta(\epsilon) \ln\left\{1 - \exp\left[-\frac{\epsilon m_p c^2}{2ET(z)}\right]\right\}^{-1}. \qquad (1)$$

Here $T$ is the MBR temperature (in energy units), $\epsilon$ is the energy of an MBR photon in the proton

– 3 –

rest frame, $\epsilon_{th}$ is the threshold energy for the considered interaction, $\sigma$ is the interaction cross section and $\eta$ is the average relative energy loss in the MBR frame. Since $T \propto (1+z)$, $\tau$ scales with redshift as $\tau(E, z) = (1+z)^{-3}\tau[(1+z)E, z=0]$. For pair production we use the results of Blumenthal (1970). For pion production we use the recent compilation of cross sections given by Hikasa et al. (1992), and $\eta(\epsilon) = (\epsilon + m_\pi^2 c^2/2m_p)/(m_p c^2 + 2\epsilon)$, obtained under the assumption [shown to be adequate in Yoshida & Teshima (1993) and in Aharonian & Cronin (1994)], that the total cross section corresponds to an isotropic production of a single pion. We approximate the resulting energy loss time due to pion production by

$$\tau_\pi(E, z=0) = \max[\tau_0 \exp(E_c/E), \tau_\infty], \tag{2}$$

with $\tau_0 = 2.33 \times 10^7$yr, $E_c = 3.25 \times 10^{20}$eV and $\tau_\infty = 4.55 \times 10^7$yr. These values give an approximation for (1) with $\sim 10\%$ accuracy for $E > 5 \times 10^{19}$eV. Berezinski & Grigor'eva (1988) have shown that for $E \ll 10^{20}$eV $\tau_\pi$ is given by

$$\tau_\pi(E, z=0) = \frac{\pi^2 c^4 \hbar^3 m_p}{2T^3(z=0)\epsilon_{th}^2} \left(\frac{d\sigma}{d\epsilon}\right)_{\epsilon=\epsilon_{th}}^{-1} \exp\left[\frac{\epsilon_{th} m_p c^2}{2T(z=0)E}\right]. \tag{3}$$

It should be noted, that we do not use the value implied by (3) for $\tau_0$. Using this value gives an accurate $\tau_\pi$ for $E \ll 10^{20}$eV, but results in a large deviation from the accurate $\tau_\pi$ above $5 \times 10^{19}$eV ($\sim 30\%$ at $10^{20}$eV). Since the energy loss due to pion production dominates over that due to pair production only above $5 \times 10^{19}$eV, we have chosen a value of $\tau_0$ which gives a better approximation at this energy range.

Using the above rates of energy loss, we compute the energy $E_0(E, z)$ at which a proton should be produced at an epoch $z$ in order to be observed at present ($z=0$) with energy $E$. We denote by $f_0(E)$ the present CR production rate per unit energy and volume, and allow a redshift evolution of the form $f(E, z) = (1+z)^{3+m} f_0(E)$ ($m = 0$ implies no evolution). The present number density per unit energy of high energy protons is then given by

$$\frac{dn}{dE} = H_0^{-1} f_0(E) \int_0^{z_{max}} dz\, g(z)(1+z)^{m-5/2} \left[\frac{E_0(E, z)}{E}\right]^{-\alpha} \frac{\partial E_0(E, z)}{\partial E}. \tag{4}$$

Here, $H_0$ is the Hubble constant and $z_{max}$ is the maximum redshift at which CR sources exist. The function $g(z) \equiv -H_0(1+z)^{5/2}(dz/dt)^{-1}$ depends on the average universe density and on the cosmological constant. For a flat universe with zero cosmological constant $g(z) \equiv 1$.

## 3. Results.

Fig. 1 presents the integral CR flux, derived from (4), produced by a homogeneous distribution of non-evolving ($m = 0$) sources with $\alpha = 2.2$. For this calculation we have used a flat universe with zero cosmological constant, $H_0 = 75$km s$^{-1}$Mpc$^{-1}$, and $z_{max} = 2$. The monocular Fly's Eye data and the AGASA data are also presented. The AGASA flux at $3 - 10 \times 10^{18}$eV is $\sim 1.7$ times higher than that reported by the Fly's Eye, corresponding to a systematic $\sim 20\%$ larger estimate of event energies in the AGASA experiment compared to the Fly's Eye experiment (see also Hayashida *et al.* 1994, Yoshida *et al.* 1995). We have therefore multiplied in Fig. 1 the Fly's Eye energy by 1.1 and the AGASA energy by 0.9. Bird *et al.* (1994) find that the Fly's Eye flux in the energy range $4 \times 10^{17} - 4 \times 10^{19}$eV can be fitted by a sum of two power laws: A steeper Galactic component with $J \propto E^{-2.5}$ dominating at lower energy, and a shallower extra-Galactic component with $J \propto E^{-1.6}$ dominating at higher energy. The Fly's Eye composition data is consistent with the Galactic component being composed of iron, and the extra-Galactic component being purely protonic. The Bird *et al.* fit to the extra-Galactic component is also shown in Fig. 1.

Let us first consider the energy range $2 \times 10^{19}$eV $\leq E < 10^{20}$eV. The data in this range is consistent with the cosmological model. For the binned data given in Bird *et al.* (1994) and Yoshida *et al.* (1995), the model gives $\chi^2 = 0.65$ per degree of freedom. Furthermore, the flux predicted by the cosmological model for $E < 2 \times 10^{19}$eV is consistent with the Bird *et al.* fit to the extra-Galactic component. The number of events observed above $2 \times 10^{19}$eV does not allow an accurate determination of $\alpha$. The value of $\alpha$ should be $< 2.8$, since for larger values the flux predicted from the model (with normalization chosen to fit the data in the range $2 \times 10^{19}$eV $\leq E < 10^{20}$eV) exceeds the observed flux at $10^{19}$eV. However, $\chi^2$ of less than 1 per degree of freedom is obtained for $1.8 < \alpha < 2.8$. For these models, the present rate of energy produced as $10^{19} - 10^{21}$eV protons is $4.5 \pm 1.5 \times 10^{44}$erg Mpc$^{-3}$ yr$^{-1}$. The systematic uncertainty in the experimental event energies, which is of order 30%, does not influence the above conclusions significantly.

We now turn to the two events observed at $> 10^{20}$eV. As seen in Fig. 1, the flux deduced from these events is higher than that predicted from the cosmological model (normalized to fit the data in the range $2 \times 10^{19}$eV $\leq E < 10^{20}$eV). It seems, that a homogeneous cosmological distribution of power law CR sources producing the CR flux in the range $2 \times 10^{19}$eV $\leq E < 10^{20}$eV, can not account for the existence of the $> 10^{20}$eV events. However, it should be noted that the statistical significance of the apparent discrepancy is not high. For the Fly's Eye exposure, the model predicts an average of $\sim 1.3$ events above $10^{20}$eV, and the probability that the first event observed at this energy range is above $2 \times 10^{20}$eV is $\sim 15\%$. For the AGASA exposure, the probability to observe an event above $10^{20}$eV is $\sim 20\%$. With 1 event observed by each experiment, the possibility that these events are produced by a homogeneous cosmological source distribution can not be ruled out. It should further be noted, that the flux above $10^{20}$eV is sensitive to inhomogeneities in the production of protons over scales of order 10 Mpc. In Fig. 2 the fraction of the integral flux contributed by sources located further than a certain distance is plotted for several distances. The flux above $10^{20}$eV is dominated by sources closer than 30 Mpc and is therefore sensitive to $\sim 10$ Mpc scale inhomogeneities. Such inhomogeneities are likely to exist, since they may arise from large scale clustering or, if the typical number of sources in a $\sim 10$ Mpc sphere is not large, from fluctuations in individual source intensity. Thus, the characteristic signature of a cosmological model, due to interaction of protons with the MBR, may be obscured above $10^{20}$eV by inhomogeneities. A local over-density of CR sources, for example, would result in a flux above $10^{20}$eV which is higher than that predicted from a cosmological source distribution homogeneous down to $\sim 10$Mpc scale.

As seen in Fig. 2, the flux above $5 \times 10^{19}$eV is dominated by sources at distances $> 100$ Mpc, and is therefore not expected to be sensitive to inhomogeneities. For this reason, the "black-body cutoff" at $\sim 5 \times 10^{19}$eV should be a robust signature of the cosmological model. It is clear from Fig. 1, that the number of events detected above $5 \times 10^{19}$eV is smaller than would be expected from a power-law extrapolation of the flux at lower energy. To estimate the significance of this "cutoff" we fitted a power law to the Fly's Eye and AGASA flux in the range $1 - 5 \times 10^{19}$eV. The minimum in $\chi^2$ (0.8 per degree of freedom) is obtained for $J \propto E^{-1.86}$, with 20.4 events expected above $5 \times 10^{19}$eV for the combined exposure of both experiments. With 12 events actually observed, the "cutoff" is detected with only $1.8\sigma$ significance. In order to rule out the "power law" hypothesis, that the flux in the range $5 - 10 \times 10^{19}$eV is described by an extrapolation of the lower energy power-law fit, larger exposures are required. If the high energy CR flux is indeed described by the cosmological model presented in Fig. 1, then the probability that the "power law" hypothesis would be ruled out with a $3\sigma$ significance increases with increased exposure. With current experiments (AGASA and Fly's Eye), additional $\sim 30$ years of observation are required for this probability to exceed 70% (For the estimate of the required time we have taken into account the fact that the AGASA experiment triggering was recently improved). The required observation time would be reduced if new, larger, CR experiments become operative: $\sim 10$ observation-years would be required with the new High Resolution Fly's Eye experiment (Corbató et al. 1992), which is planned to become operative in two years; Only $\sim 1$ year of observation would be required if the proposed $\sim 5000$ km$^2$ detectors are built (Cronin 1992, Watson 1993).

The flux above $10^{19}$eV is produced in a cosmological model by sources at distances smaller than 1Gpc (see Fig. 1). For this reason, the flux predicted is insensitive to the assumed values of the average universe density and cosmological constant, and also insensitive to source evolution and to the maximal redshift $z_{max}$ (as long as $z_{max} > 0.5$). The flux does depend on the value of the Hubble constant. However, for the reasonable range of $50 - 100$ km s$^{-1}$Mpc$^{-1}$, the changes in the predicted flux are small ($< 10\%$ at a fixed energy).

## 4. Conclusions

We have shown that the CR spectrum in the range $10^{19}$eV $< E < 10^{20}$eV, reported by the Fly's Eye and the AGASA experiments, is consistent with that expected from a homogeneous cosmological distribution of CR sources, each generating high energy protons with a power law differential spectrum $dN/dE \propto E^{-\alpha}$. With the small number of events observed, the value of $\alpha$ is only constrained to the range $1.8 < \alpha < 2.8$. The rate of energy produced by the cosmological sources as $10^{19} - 10^{21}$eV protons should be $4.5 \pm 1.5 \times 10^{44}$erg Mpc$^{-3}$ yr$^{-1}$. As recently pointed out by Waxman (1995), this rate is comparable to that produced in $\gamma$-rays by cosmological $\gamma$-ray

– 6 –

bursts. The "black-body cutoff", expected in a cosmological model at $\sim 5 \times 10^{19}$eV, is observed in current data with only marginal statistical significance. Raising the statistical significance to a $\sim 3\sigma$ level would require $\sim 30$ years of observations with current experiments [$\sim 10$ years with the new High Resolution Fly's Eye experiment, planned to become operative in two years (Corbató *et al.* 1992)], but only $\sim 1$ year if the proposed $\sim 5000$ km$^2$ detectors (Cronin 1992, Watson 1993) are built. The flux just above $5 \times 10^{19}$eV is dominated by sources at distances $> 100$ Mpc, and is not expected to be sensitive to source inhomogeneities. This implies that the "black-body cutoff" is a robust signature of the cosmological model. It also implies that no anisotropy related to large scale source clustering is expected in the angular distribution of cosmic ray events with energies up to $\sim 5 \times 10^{19}$eV.

The flux above $10^{20}$eV is dominated by sources at distances $< 30$ Mpc, and is therefore likely to be sensitive to source inhomogeneities. Therefore, above $10^{20}$eV inhomogeneities may obscure the characteristic signature of a cosmological model, which is produced by the interaction of protons with the MBR. Nevertheless, the two events measured by the Fly's Eye and the AGASA experiments are not inconsistent with the homogeneous cosmological model, which predicts an average of one event above $10^{20}$eV for the Fly's Eye exposure. The detection of $> 10^{20}$eV events is, of course, extremely important, since they are likely to point to the location of nearby extra-Galactic high energy CR sources.

I thank J. N. Bahcall for invaluable suggestions and comments, and the unknown referee for constructive criticism which led to an improved manuscript. This research was partially supported by a W. M. Keck Foundation grant and NSF grant PHY 92-45317.

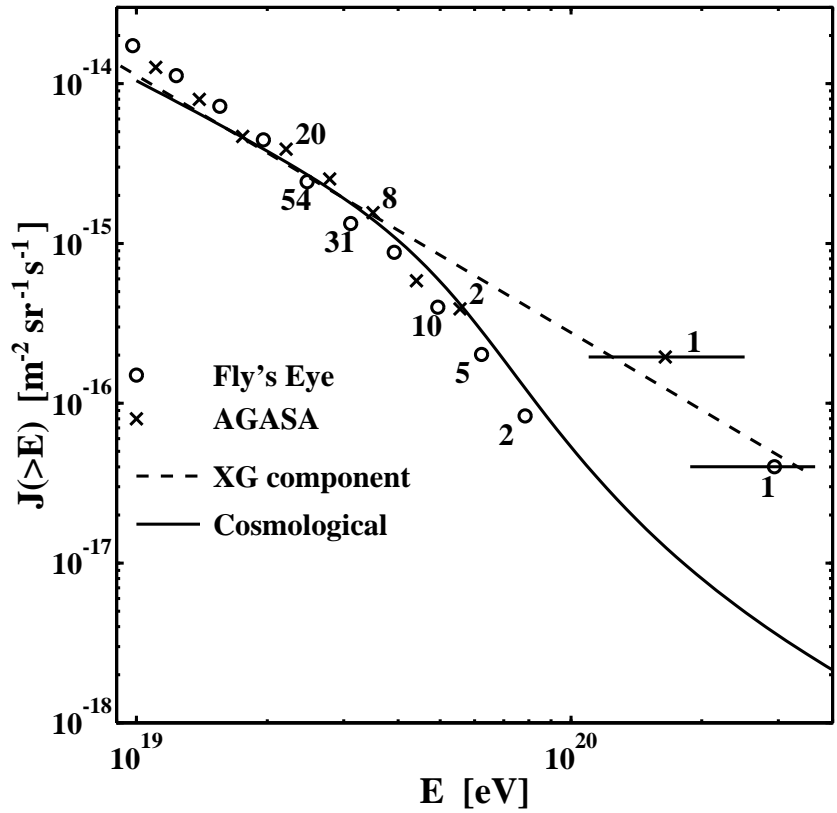

Fig. 1.— The integral CR flux expected from a cosmological distribution of sources with $\alpha = 2.2$, compared to the Fly's Eye and AGASA data. The integers shown in the figure denote the number of events observed. The flux deduced from the highest energy events is plotted with the $2\sigma$ error bars for the event energy. The dashed line denotes the fit by Bird *et al.* (1994) for the extra-galactic flux.

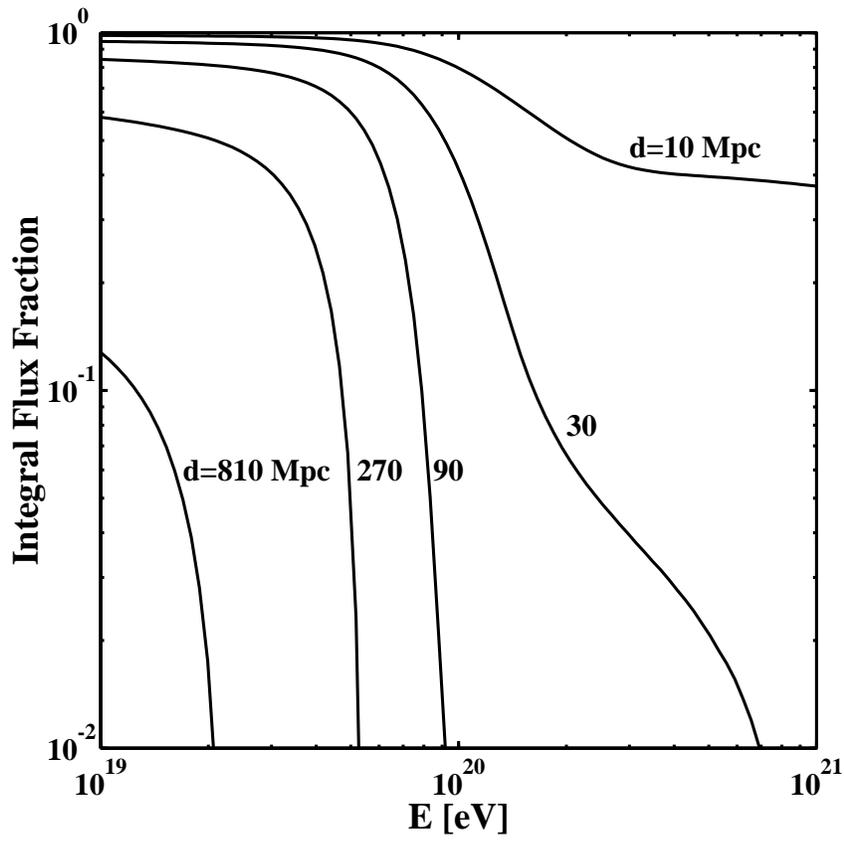

Fig. 2.— The fraction of the integral flux contributed by sources more distant than $d$.